\documentclass[prl,twocolumn,showpacs,superscriptaddress]{revtex4-1}

\usepackage{graphicx}
\usepackage{amsmath,amssymb,bm}

\usepackage{color}

\begin{document}


\title{
$1/(N-1)$ expansion based on a perturbation theory in $U$
\\
for the Anderson model with $N$-fold degeneracy
}


\author{A.\ Oguri}
\affiliation{%
Department of Physics, Osaka City University, Sumiyoshi-ku, 
Osaka, Japan\\
}

\author{R.\ Sakano}
\affiliation{%
Department of Applied Physics, University of Tokyo, Bunkyo, Tokyo, Japan\\
}

\author{T.\ Fujii}%
\affiliation{%
Institute for Solid State Physics, University of Tokyo, Kashiwa, Chiba, Japan\\
}%


\date{\today}

\begin{abstract}
We study low-energy properties of the $N$-fold degenerate Anderson model.
Using a scaling that takes $u=(N-1) U$ as an independent variable 
in place of the  Coulomb interaction $U$,
the perturbation series in $U$ is reorganized 
as an expansion in powers of $1/(N-1)$.
We calculate the renormalized parameters, 
which characterize the Kondo state, to the next leading order 
in the $1/(N-1)$ expansion at half-filling.
The results,  especially the Wilson ratio, 
agree very closely with the exact numerical 
renormalization group results at $N=4$.  
This ensures the applicability of our approach 
to $N > 4$, and 
we present highly reliable results 
for nonequilibrium Kondo transport through a quantum dot. 
\end{abstract}

\maketitle


The Anderson impurity has been studied extensively 
as a model for strongly correlated electrons 
in dilute magnetic alloys, quantum dots,  
and also for  bulk 
systems in conjunction 
with the dynamical mean-field theory \cite{Hewson_book}. 
For quantum dots, the nonequilibrium Kondo effect can occur 
when a bias voltage is applied between two leads. 
A universal Fermi-liquid behavior \cite{Nozieres,YY2,ZlaticHorvatic,Yoshimori} 
has been closely examined at low energies for  
the steady current    
\cite{Grobis,ScottNatelson,KNG,ao2001,FujiiUeda,HBA} and 
shot noise \cite{Delattre,GogolinKomnik,Golub,Sela2009,Mora2009,Fujii2010}.

Orbital degeneracy in the impurity states  
also affects the nonequilibrium properties 
at low energies.
Recently, Mora {\it et al.\ }\cite{Mora2009}  have succeeded 
to express the current noise 
in terms of the Fermi-liquid parameters 
\cite{Nozieres,YY2,ZlaticHorvatic,Yoshimori} 
in an SU($N$) Kondo regime, 
where the Coulomb repulsion $U$ is so large that 
charge fluctuations are suppressed 
near the impurity with $N$-fold degeneracy. 
A complemental expression that 
takes into account the fluctuations at half-filling  
has been presented in our previous work \cite{Sakano}.
In this case, the corrections due to finite $U$ enter 
through the Wilson ratio $R$, which is a correlation function 
defined with respect to the equilibrium ground state, 
and through the width of the Kondo resonance $\widetilde{\Delta}$.
Therefore, explicit values of these two parameters, 
$R$ and $\widetilde{\Delta}$, 
are required to study the low-energy transport thoroughly.
The exact numerical renormalization group (NRG) approach 
is still applicable to multi-orbital systems. 
It practically works, however, for small degeneracies 
$N\leq 4$ \cite{Sakano,Nishikawa1}, 
which for $N=2$ corresponds to the spin degeneracy. 
Therefore, alternative approaches are needed 
to explore the large degeneracies at $N>4$.

In this Letter, we propose a systematic approach to 
calculate correlation functions at $N > 4$,
using a scaling that takes $u =(N-1)U$ 
as an independent variable in place of $U$.
Here, the factor $N-1$ corresponds to  
the number of different impurity states, 
with which a local electron in the impurity site can interact. 
With this scaling, the perturbation series in $U$ can be 
reorganized as an expansion in powers of $1/(N-1)$, 
 using a diagrammatic classification similar 
to the one for the $N$-component $\varphi^4$ model \cite{WilsonKogut}.
However, our approach is completely different from  
the usual $1/N$ expansion 
and non-crossing approximation,  
which are constructed 
on the basis of the perturbation expansion in the hybridization matrix 
element $v_{\nu}$ \cite{Bickers,Haule,OtsukiKuramoto}.
We calculate $R$ and $\widetilde{\Delta}$
up to the next leading order terms in the $1/(N-1)$ expansion 
at half-filling, and find that
the results agree very closely with the NRG results at $N=4$, 
 where $N$ is still not so large.
Particularly, the Wilson ratio shows 
an excellent agreement over the whole range of $U$.
The early convergence of the expansion implies 
that our scaling procedure efficiently captures the orbital effects,
and ensures the applicability to $N>4$.
This enables us to present highly reliable results 
for the nonequilibrium steady current and shot noise for $N>4$.
Our approach could have wide application to quantum 
impurities, and 
 could be used as a solver for the dynamical mean-field theory \cite{DMFT}.


The Hamiltonian for the $N$-fold degenerate Anderson model 
connected to two leads ($\nu=L,\,R$) is given by 
\begin{align}
&
{\cal H} \,= \,  {\cal H}_0 + {\cal H}_U, 
 \qquad  
{\cal H}_U 
=  \,  \frac{1}{2} \sum_{m \neq m'} U \, 
n_{dm}^{} n_{dm'}^{} ,
\label{eq:mia-model} 
\\
&\!
{\cal H}_0 =  
\sum_{\nu=L,R}
\sum_{m=1}^N 
\int_{-D}^D  \! d\epsilon\,  \epsilon\, 
 c^{\dagger}_{\epsilon \nu m} c_{\epsilon \nu m}^{}
+ \sum_{m=1}^N 
\epsilon_{d}^{} 
 \,d_{m}^{\dagger} d_{m}^{} ,
 \nonumber \\
&  \quad \    + 
\sum_{\nu=L,R}\,
\sum_{m=1}^N 
 v_{\nu}^{} \left(
d_{m}^{\dagger} 
\psi^{}_{\nu m} 
+ \mbox{H.c.} \right) .
\label{eq:AndersonHamiltonian}
\end{align}
Here, $d_m^{\dagger}$ creates an electron 
with energy $\epsilon_{d}$ in orbital $m$ at the impurity site, 
$n_{dm}^{} = d_{m}^{\dagger} d_{m}^{}$, and 
 $m$ ($=1,2, \cdots, N$) includes the spin degrees of freedom. 
 $c_{\epsilon\nu m}^{\dagger}$ creates 
a conduction electron with energy $\epsilon$ and orbital $m$ 
in lead $\nu$, and is normalized as 
$
\{ c^{\phantom{\dagger}}_{\epsilon\nu m}, 
c^{\dagger}_{\epsilon'\nu'm'}
\} = \delta_{\nu\nu'} \,\delta_{mm'}   
\delta(\epsilon-\epsilon')$. The linear combination 
$\psi^{}_{\nu m} \equiv  \int_{-D}^D d\epsilon \sqrt{\rho} 
\, c^{\phantom{\dagger}}_{\epsilon\nu m}$,
with $\rho=1/(2D)$, couples to the impurity level via 
the hybridization matrix element $v_{\nu}^{}$, 
and $\Delta \equiv \Gamma_L + \Gamma_R$ with $\Gamma_{\nu} = 
\pi \rho\, v_{\nu}^2$. 
We consider the parameter region where 
$\Delta$, $\epsilon_d$, and $U$ are  
much smaller than the half band width $D$.

We use the imaginary-frequency Green's function that takes the form
$G(i\omega) 
=  \left[
i\omega - \epsilon_d^{} 
+ i \Delta \, \mathrm{sgn}\,\omega 
- \Sigma (i\omega)\right]^{-1}
$ for $|\omega| \ll D$.
The behavior of the self-energy $\Sigma(i\omega)$ for  
small $\omega$ determines 
the enhancement factor for the linear specific heat 
$\widetilde{\gamma} =  
1 -  {\partial\Sigma(i\omega)}/{\partial (i\omega)}
|_{\omega=0}$, and 
the renormalized parameters 
$z= {1}/{\widetilde{\gamma}}$, 
$\,\widetilde{\epsilon}_{d} =   z[ \epsilon_{d} + \Sigma (0)]$, 
and 
$\widetilde{\Delta}  = z \Delta$. 
The average number of local electrons  
can be deduced from the phase shift
$\delta \equiv   
\cot^{-1} ({\widetilde{\epsilon}_d^{}}/{\widetilde{\Delta}})$, 
using the Friedel sum rule, 
$\langle n_{dm} \rangle = \delta/\pi$. 
The enhancement factor for the spin 
susceptibility and that for the charge  
can be written in the form 
$\widetilde{\chi}_{s}^{} \equiv
\widetilde{\chi}_{mm}^{} - 
\widetilde{\chi}_{mm'}^{}$ and 
$\widetilde{\chi}_{c}^{}  \equiv 
\widetilde{\chi}_{mm}^{} + (N-1)\, \widetilde{\chi}_{mm'}^{}$ 
for $m\neq m'$.
These susceptibilities can be deduced from 
 the self-energy and  four-point vertex function  
$\Gamma_{mm';m'm}^{}(i \omega_1,i\omega_2;i\omega_3,i\omega_4)$ 
for $m\neq m'$, using the Ward identities \cite{Yoshimori},  
\begin{align}
\widetilde{\chi}_{mm}^{} = \widetilde{\gamma} , 
\quad \ \ 
\widetilde{\chi}_{mm'}^{} =    -\,
\frac{\sin^2 \delta}{\pi\Delta}
\, \Gamma_{mm';m'm}^{}(0,0;0,0) .
\label{eq:ward_fermi}
\end{align}
Furthermore, 
$\widetilde{U} \equiv z^2 \Gamma_{mm';m'm}^{}(0,0;0,0)$ 
corresponds to the residual interaction between the quasi-particles.

The Wilson ratio $R$ parameterizes how far the system is away 
from the Kondo limit, and plays a central role for finite $U$, 
\begin{align}
R  \equiv \frac{\widetilde{\chi}_{s}}{\widetilde{\gamma}}
 =  1+\frac{\widetilde{g}}{N-1} \, \sin^2\delta 
,
\qquad 
\frac{\widetilde{\chi}_{c}}{\widetilde{\gamma}}
 =   1- \widetilde{g}  \sin^2\delta .
\label{eq:Wilson}
\end{align}
Here,  the scaling factor $N-1$ is introduced 
to the renormalized interaction 
$\widetilde{U}$ and the bare one $U$, 
such that 
\begin{align}
\widetilde{g} \,\equiv\,  
 (N-1)\,\frac{\widetilde{U}}{\pi \widetilde{\Delta}} 
\;, \qquad  \quad 
g \,\equiv\,  
 (N-1)\,\frac{U}{\pi\Delta} \;.
\label{eq:g_scale}
\end{align}%
In the following 
we consider the particle-hole symmetric case, 
where $\epsilon_{d}^{} = - (N-1)U/2$ and  $\delta = \pi/2$.  
In this case, the renormalized coupling takes a value in the range 
 $0\leq \widetilde{g} \leq 1$.
It approaches to $\widetilde{g} \to 1$ in the limit of $g \to \infty$ 
as the charge fluctuation is suppressed $\widetilde{\chi}_{c}\to 0$.

We calculate  $\widetilde{\gamma}$ 
and $\Gamma_{mm';m'm}^{}(0,0;0,0)$ perturbatively 
to order $U^3$ and $U^4$, respectively, by 
extending Yamada's calculations for $N=2$ \cite{YY2} 
to general $N$ \cite{Sakano}, and obtain

\begin{widetext}
\begin{align}
\widetilde{g} \,=& \ \
g - \frac{N-2}{N-1} \ g^2
+ \frac{ 
(N-1)^2-\frac{\pi^2}{4}(N-1) + (11 - \pi^2)
}{(N-1)^2} \ g^3 
\nonumber \\
& 
\quad - \, 
\frac{(N-2) 
\biggl[\, (N-1)^2 - \left( 6+ \pi^2 - \frac{21}{2} \zeta (3) \right)(N-1) 
+ \left(\frac{175}{2} \zeta (3)  
- \frac{23}{3} \pi^2
- 28\right)  
\,\biggr]
}
{(N-1)^3} \ g^4
\ + \, O(g^5)\;,
\label{eq:WR-ex2}
\\
\widetilde{\gamma} =& \ \  
1+ \frac{1}{N-1}
\left[
\,\left( 3-\frac{\pi^2}{4}\right)  g^2 
-\left( \frac{21}{2} \zeta(3) - 7 - \frac{\pi^2}{2} \right)
\frac{N-2}{N-1}\  g^3 
\  + \,  O(g^4) \, \right]
\;. 
\label{eq:gamma_tilde_in_U} 
\end{align}
\end{widetext}
Here, $\zeta(x)$ is the Riemann zeta function, which disappears 
at $N=2$ where the impurity has only 
the spin degeneracy \cite{ZlaticHorvatic}.
For $N>2$, 
 $\widetilde{\gamma}$ and $\widetilde{g}$ 
are no longer even nor odd function of $U$.
We see in Eqs.\  \eqref{eq:WR-ex2}  and \eqref{eq:gamma_tilde_in_U} 
that the coefficients in the perturbation series 
can be expanded in powers of $1/(N-1)$. 
Thus, the perturbation series in $g$ 
can be reorganized as an expansion with respect to $1/(N-1)$.
If the $N\to \infty$ limit is taken at fixed $g$,
then the right hand side of Eq.\ \eqref{eq:WR-ex2} 
approaches to an alternating geometric series in $g$,
and 
$\widetilde{\gamma}$ approaches to the noninteracting 
value $\widetilde{\gamma}\to 1$. 
We will see later that these are true for all order in $g$, 
and the asymptotic forms of 
Eqs.\  \eqref{eq:WR-ex2}  and \eqref{eq:gamma_tilde_in_U} 
in the large $N$ limit are given by
\begin{align}
\!\!
\widetilde{g} =   \frac{g}{1+g} +O\!\left(\!\frac{1}{N-1}\!\right),
\quad \ \  
\widetilde{\gamma} =   1 +O\!\left(\!\frac{1}{N-1}\!\right). \!
\label{eq:g_inf}
\end{align}
The corrections due to finite $N$ can be extracted, 
using a diagrammatic representation of the perturbation in $U$.

\begin{figure}[t]
 \leavevmode
\begin{minipage}{1\linewidth}
\includegraphics[width=0.75\linewidth]{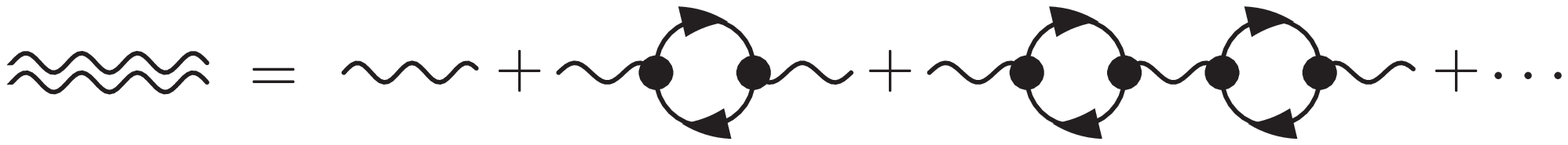}
\end{minipage}
\caption{The leading order diagrams in the $1/(N-1)$ expansion.
The wavy and solid lines indicate 
the Coulomb repulsion $U$ and 
unperturbed Green's function $G_0$, respectively.
The double wavy line represents the sum of the bubble diagrams, 
and corresponds to $\mathcal{U}_\mathrm{bub}(i\omega)$ 
given in  Eq.\  \eqref{eq:vertex_rings}. 
}
 \label{fig:vertex_rings}
\end{figure}

\begin{figure}[t]
\begin{minipage}{1\linewidth}
\includegraphics[width=0.24\linewidth]{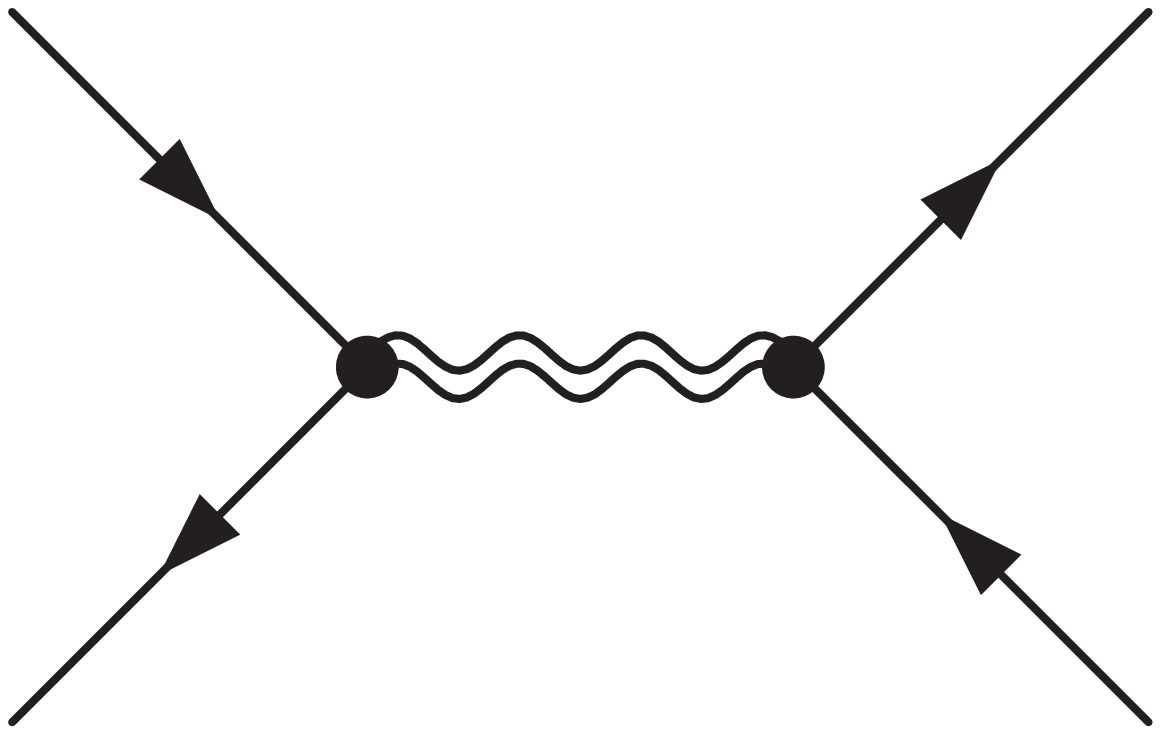}
\rule{0.12\linewidth}{0cm}
\includegraphics[width=0.35\linewidth]{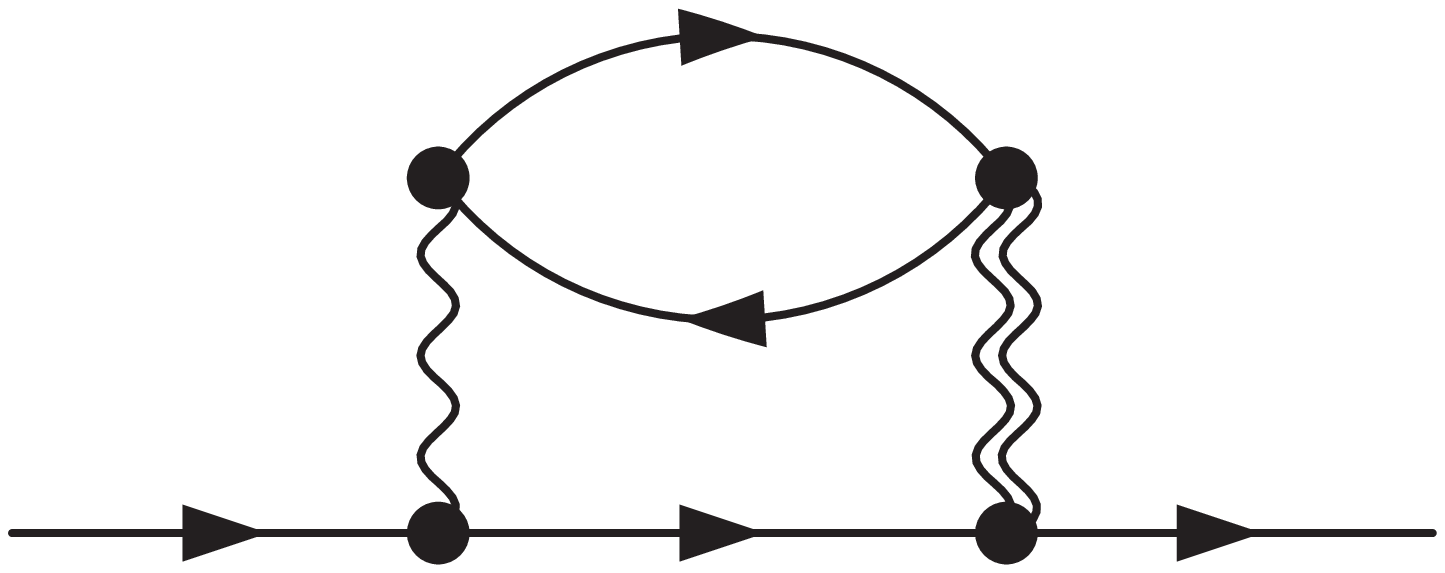}
\end{minipage}
\caption{
The diagrams which provide the order $1/(N-1)$ contributions  
with some higher order corrections [see Eq.\ \eqref{eq:vertex_rings}].
}
 \label{fig:sg_rpa}
\end{figure}

The leading order contributions in the $1/(N-1)$ expansion 
arise form a series of the bubble diagrams indicated 
in Fig.\ \ref{fig:vertex_rings}, and the sum of these diagrams corresponds to 
\begin{align}
&
 \!\! 
\mathcal{U}_\mathrm{bub}^{}(i\omega) 
=   
\frac{\phi(i\omega)}{N-1} 
+ \frac{g \pi \Delta \, \Pi(i\omega)}{(N-1)^2} 
 +  O\!\left(\!\frac{1}{(N-1)^3}\!\right) , 
\label{eq:vertex_rings}
\\
&
\phi(i\omega) 
\equiv    
\frac{g \pi \Delta}{1+g \pi \Delta \chi_0^{}(i\omega)} 
, 
\label{eq:vertex_rpa2} 
\quad  
 \Pi(i\omega)
 \equiv  
 \chi_0^{}(i\omega)\,\phi(i\omega) .
\end{align}
Here, 
$\chi_0^{}(i\omega)  \equiv    - 
\int
 \! \frac{d\omega'}{2\pi} 
G_0^{}(i\omega +i\omega') G_0^{}(i\omega')$, and    
$G_0^{}(i\omega) = [i\omega -E_d + i \Delta \, \mathrm{sgn}\,\omega]^{-1}$ with  $E_d=0$  \cite{FootNote1}. Thus 
 $\chi{_0^{}}(i\omega) =  \frac{1}{\pi \Delta} 
 \frac{2\log\left(1+|x|\right)}{|x|(2+|x|)}$ 
 with $x={\omega}/{\Delta}$.
The propagator $\mathcal{U}_\mathrm{bub}^{}(i\omega)$ 
contains not only the leading order,   
but also higher order contributions in the $1/(N-1)$ expansion.
This is because the orbital indices 
for adjacent bubbles have to be different,
and summations over internal $m$'s are {\it not\/} independent.  
The order $1/(N-1)$ contributions to  
the vertex and self-energy 
come from the diagrams shown in
 Fig.\ \ref{fig:sg_rpa}.

%
%

\begin{figure}[t]
 \leavevmode
\begin{minipage}{1\linewidth}
\includegraphics[width=0.27\linewidth]{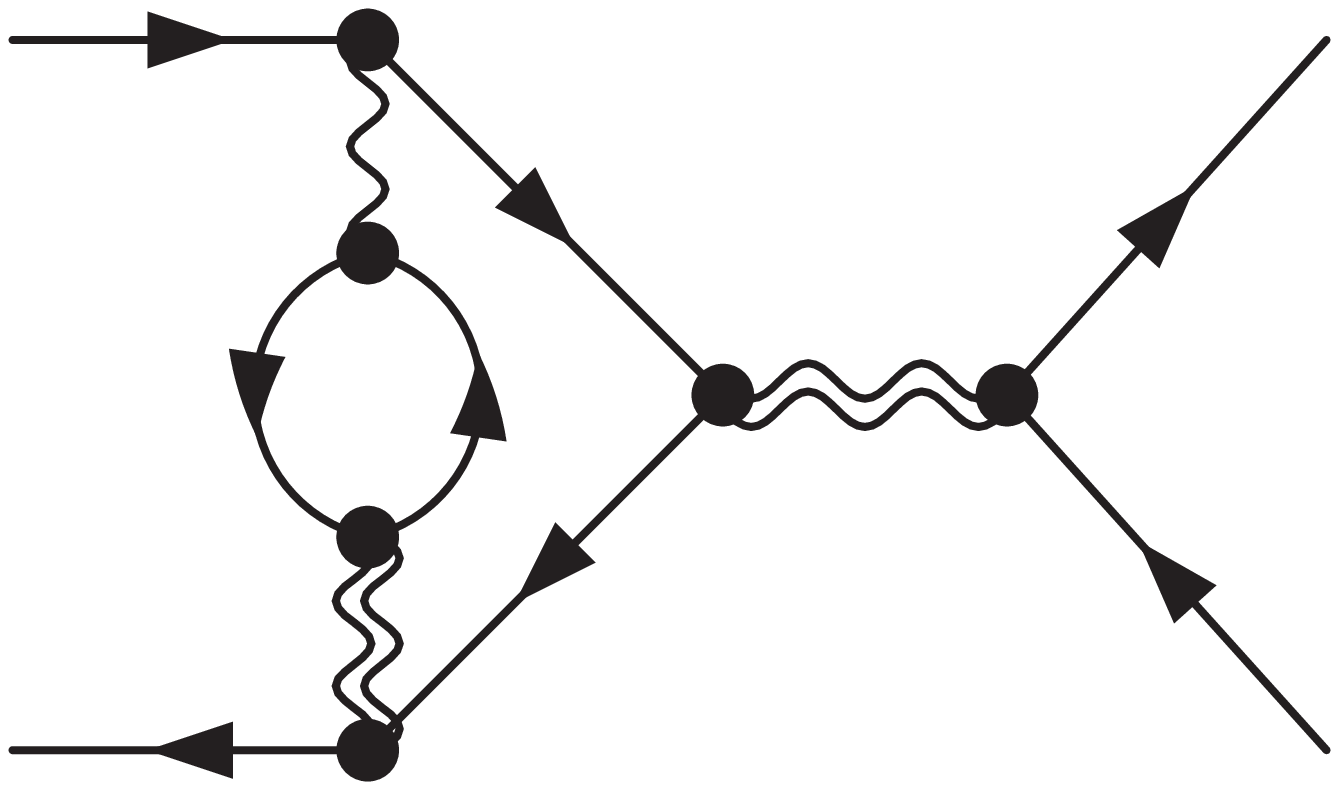}
 \rule{0.1\linewidth}{0cm}
\includegraphics[width=0.27\linewidth]{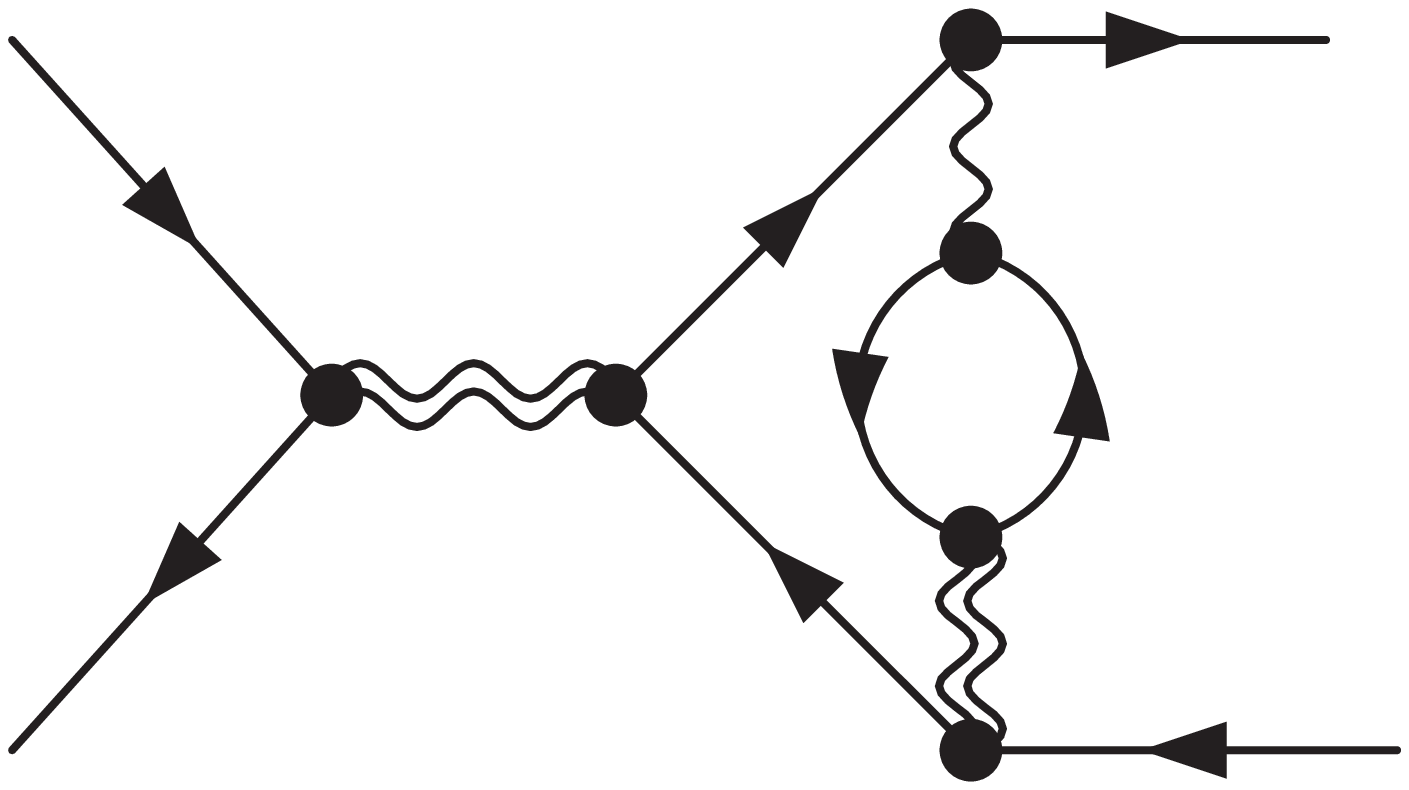}
\end{minipage}
 \rule{0cm}{0.25cm}
 \begin{minipage}{1\linewidth}
 \includegraphics[width=0.27\linewidth]{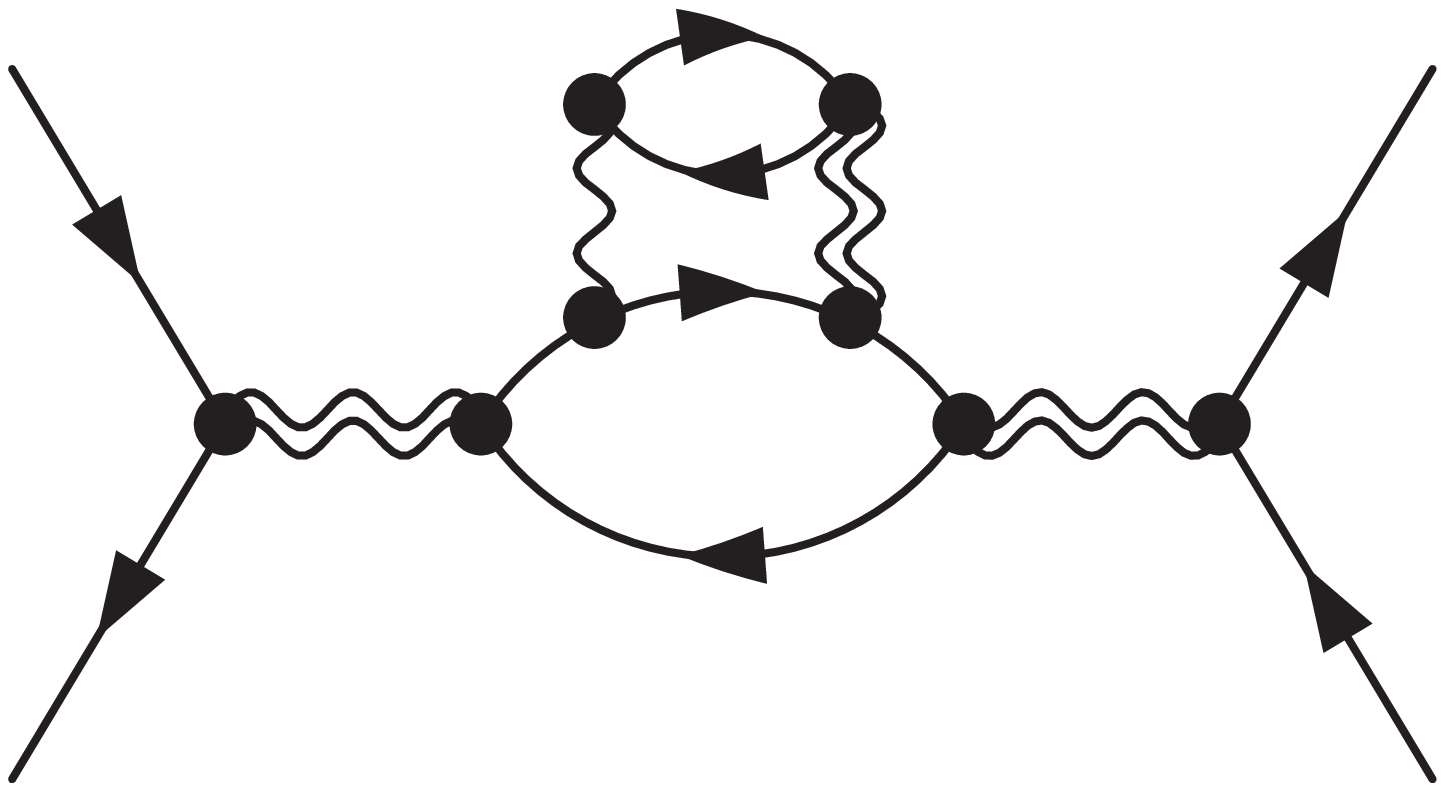}
 \rule{0.04\linewidth}{0cm}
 \includegraphics[width=0.27\linewidth]{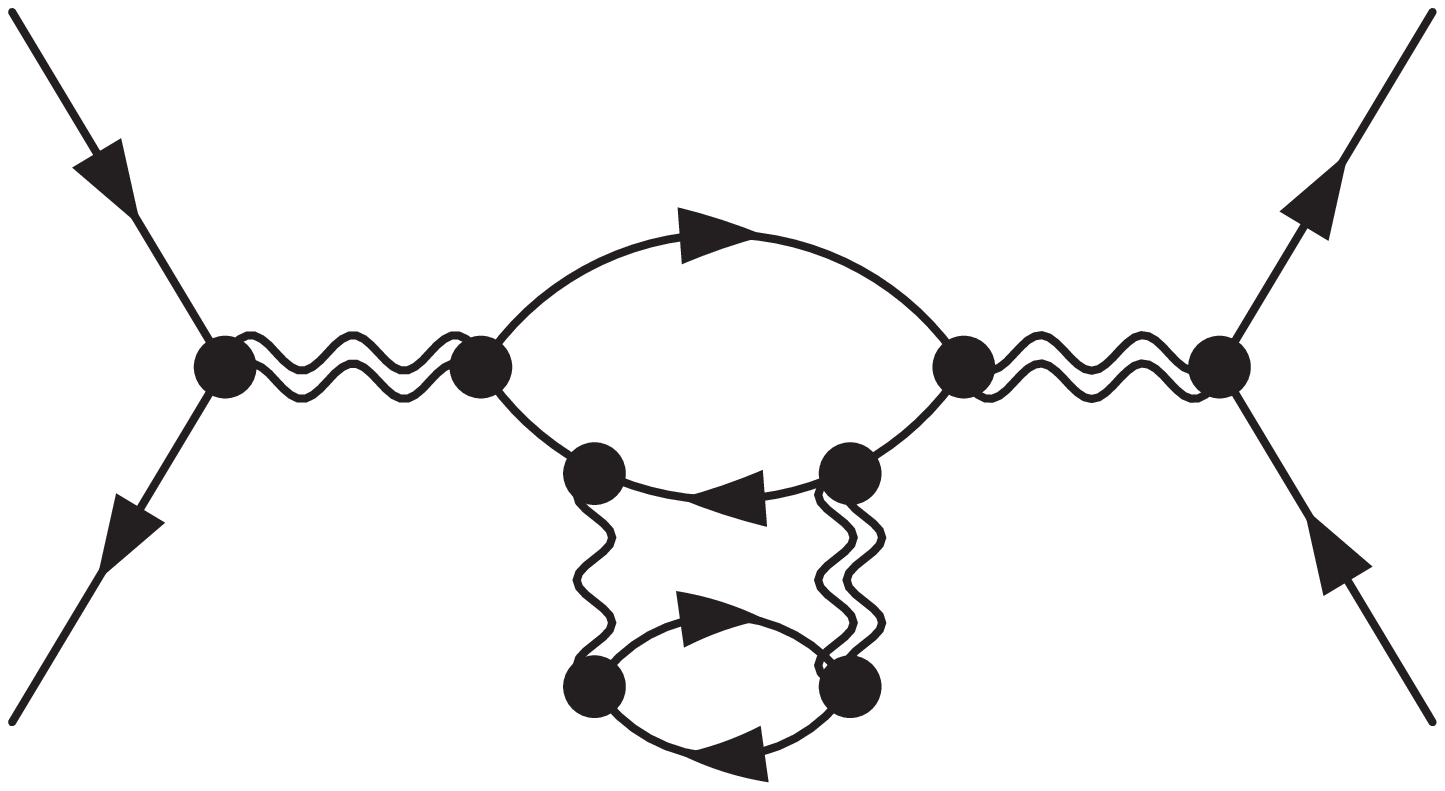}
 \rule{0.04\linewidth}{0cm}
 \includegraphics[width=0.27\linewidth]{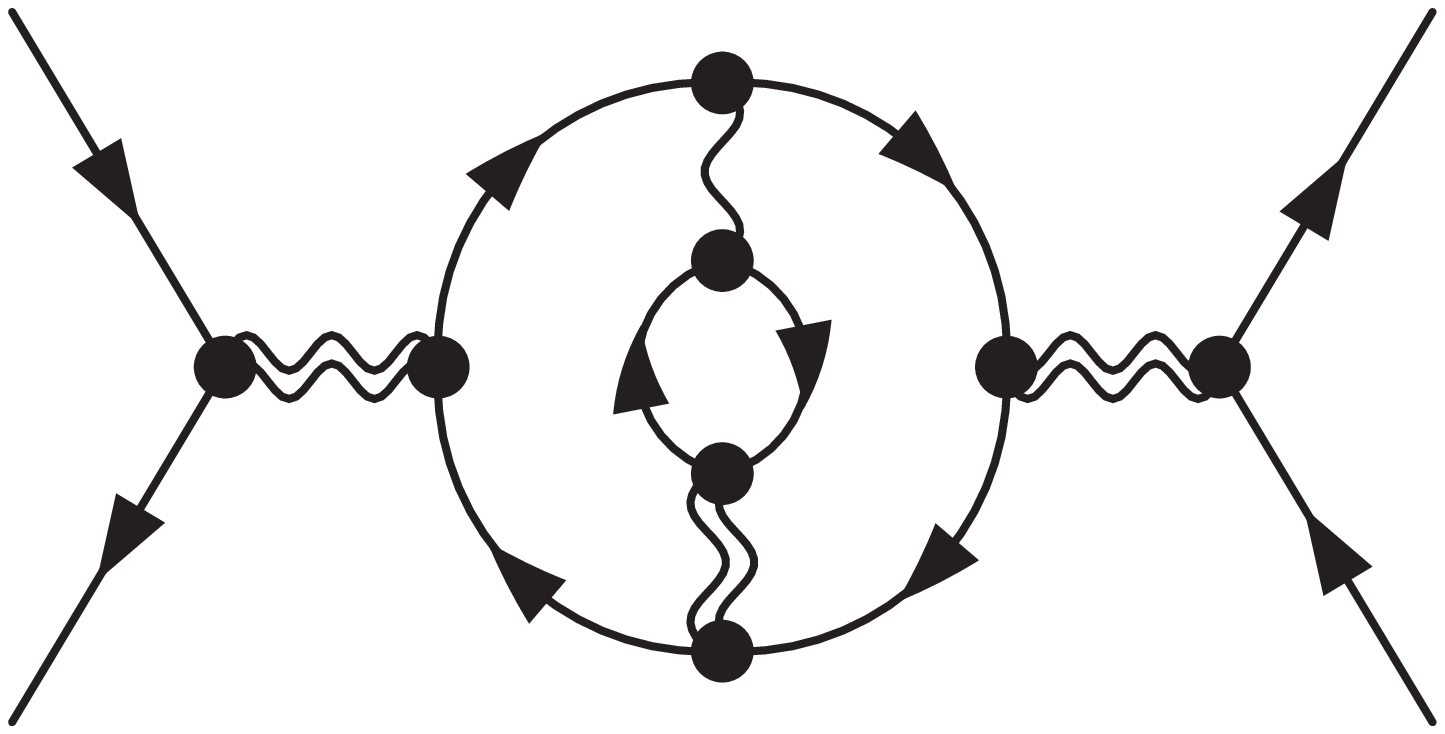}
 \end{minipage}
 \caption{
The order $1/(N-1)^2$ diagrams for the vertex function 
$\Gamma_{mm';m'm}^{}(0,0;0,0)$ for $m \neq m'$. 
}
 \label{fig:vertex_rpa}
\end{figure}

\begin{figure}[t]
 \leavevmode
\begin{minipage}{1\linewidth}
\includegraphics[width=0.24\linewidth]{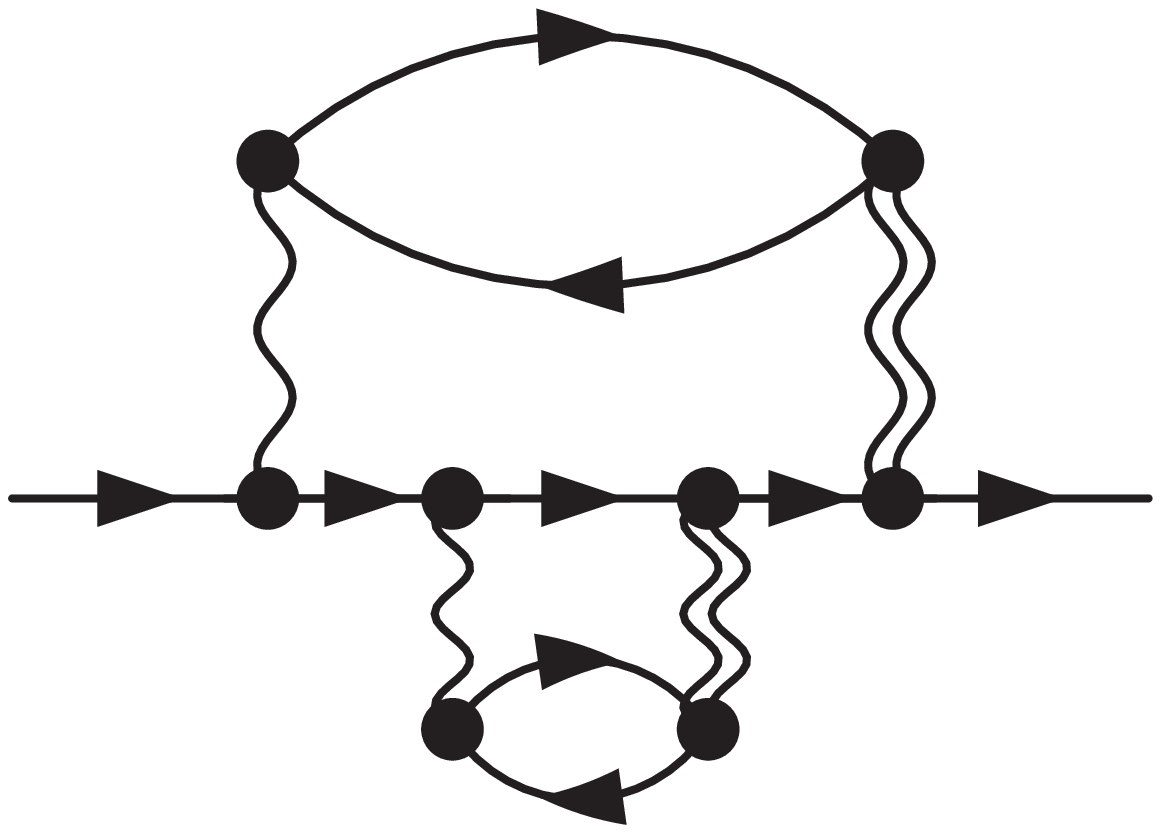}
 \rule{0.1\linewidth}{0cm}
\raisebox{-0.3cm}{
\includegraphics[width=0.26\linewidth]{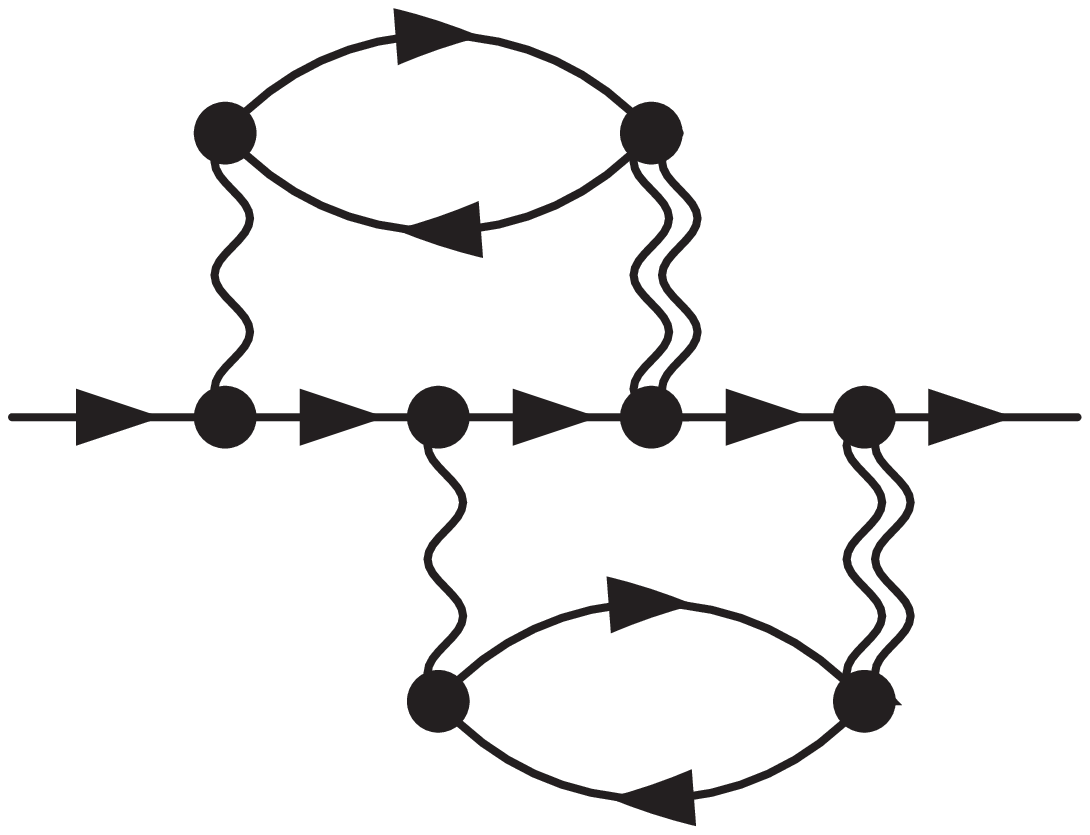}
}
\end{minipage}
 \begin{minipage}{1\linewidth}
 \includegraphics[width=0.27\linewidth]{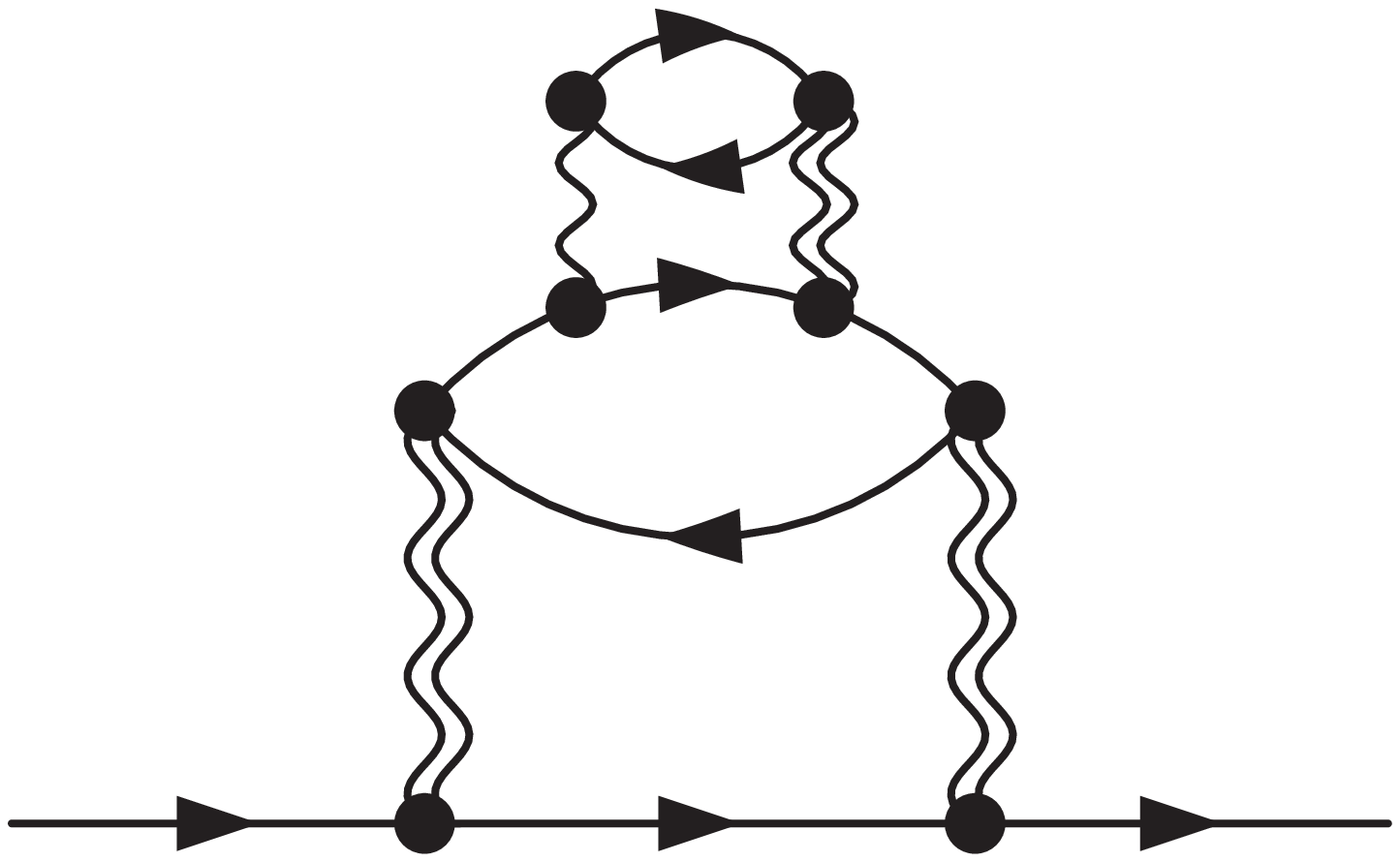}
 \rule{0.04\linewidth}{0cm}
 \includegraphics[width=0.27\linewidth]{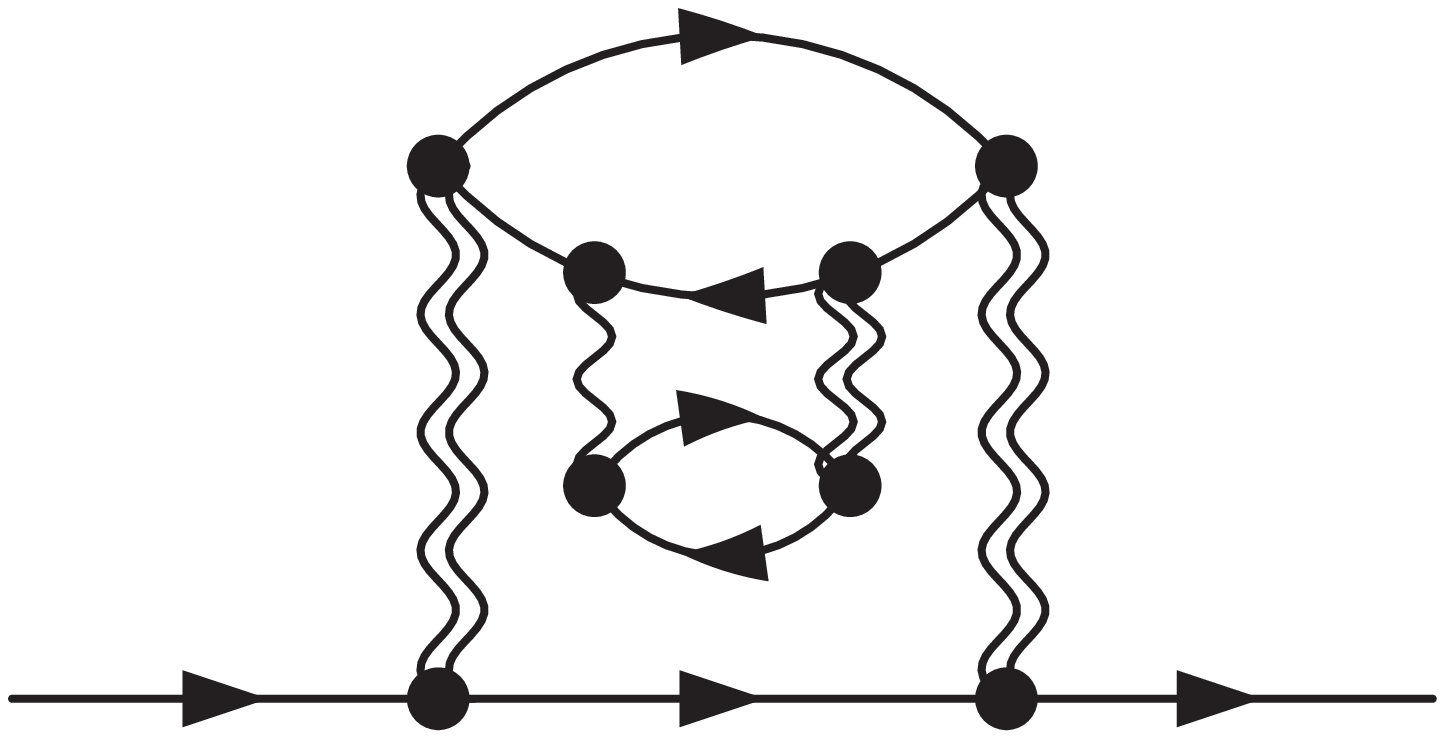}
 \rule{0.04\linewidth}{0cm}
 \includegraphics[width=0.27\linewidth]{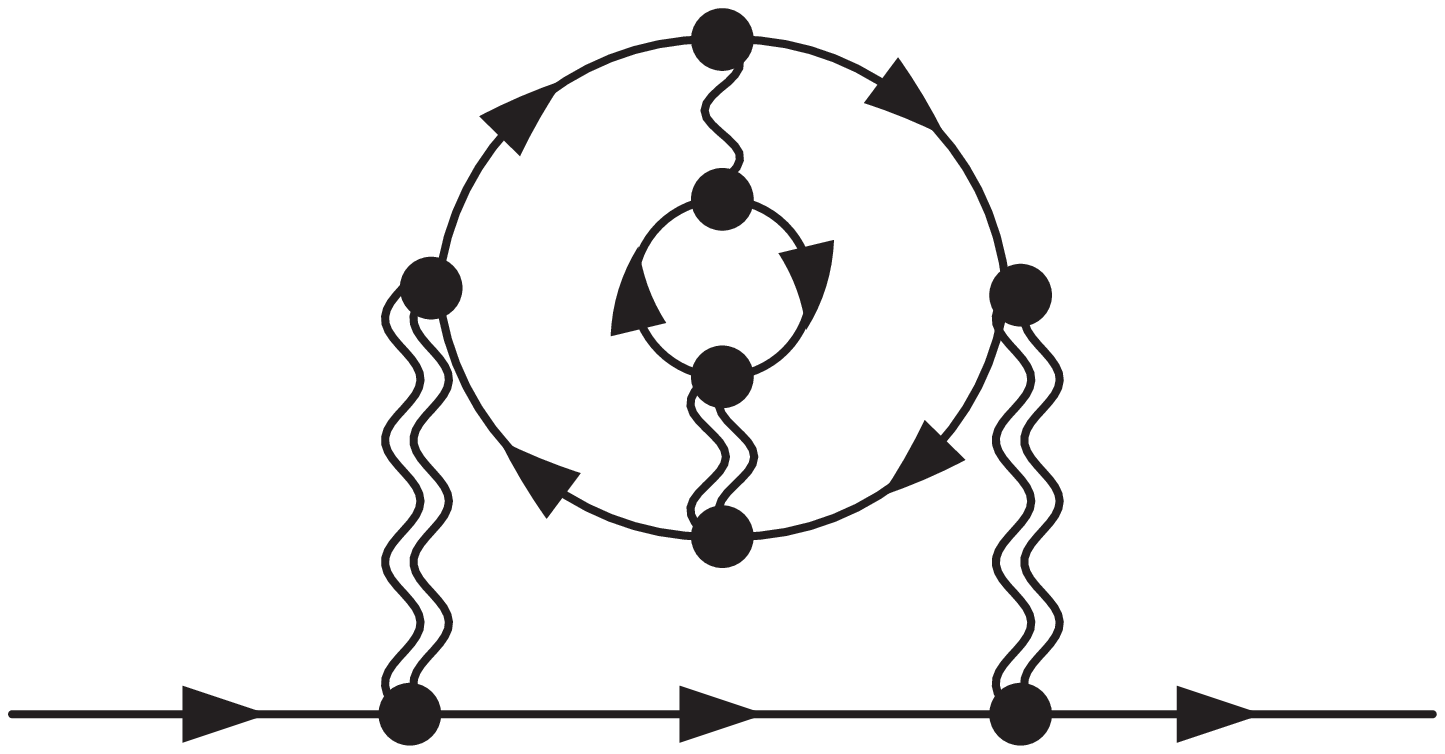}
 \end{minipage}
 \caption{
The order $1/(N-1)^2$ self-energy diagrams which contribute to 
the renormalization factor $z$ ($=1/\widetilde{\gamma}$\/).
}
\label{fig:sg_rpa_n2}
\end{figure}

\begin{figure}[t]
\leavevmode
\begin{minipage}[t]{0.48\linewidth}
\includegraphics[width=\linewidth]{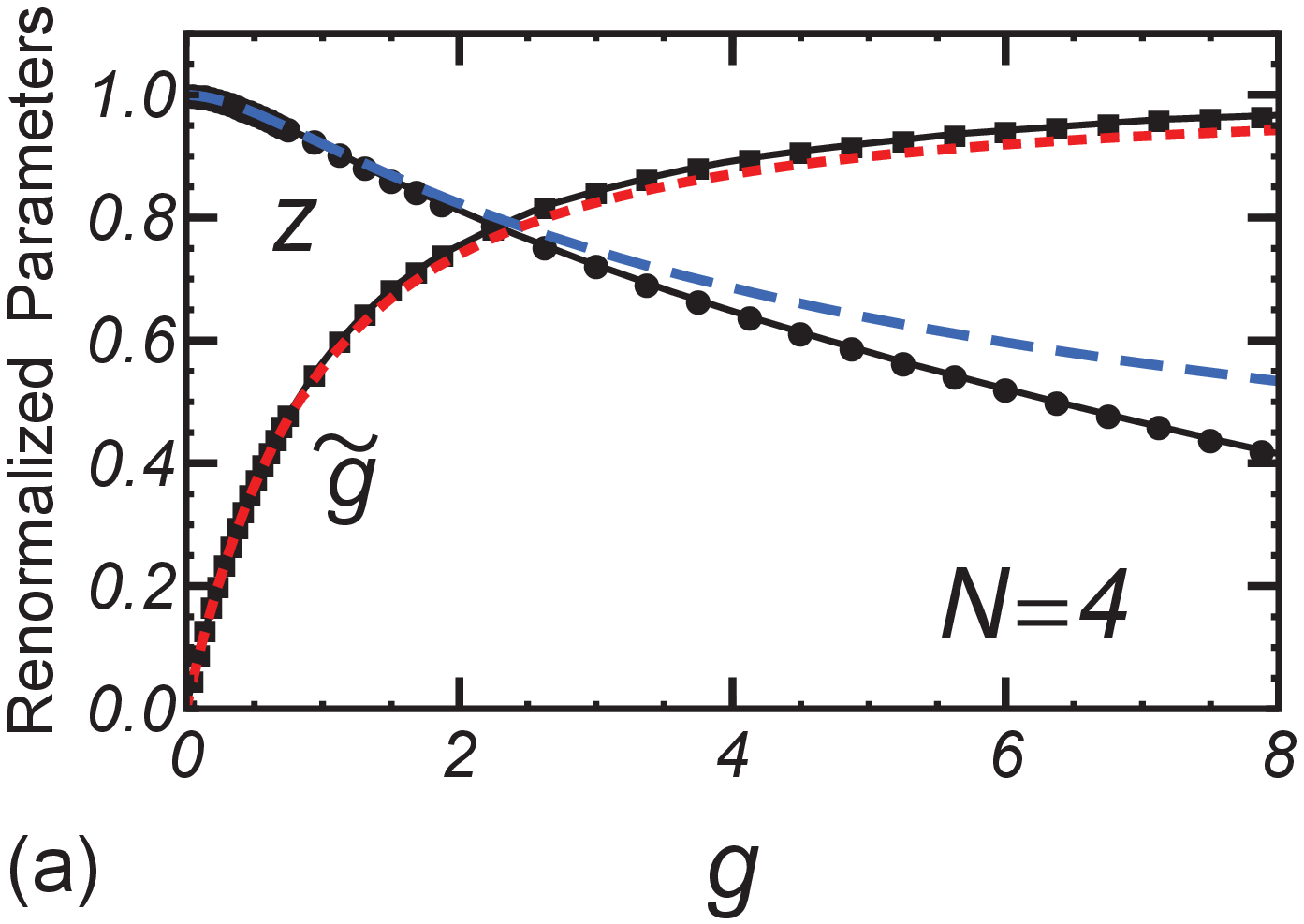}
\end{minipage}
\leavevmode
\begin{minipage}[t]{0.5\linewidth}
\includegraphics[width=\linewidth]{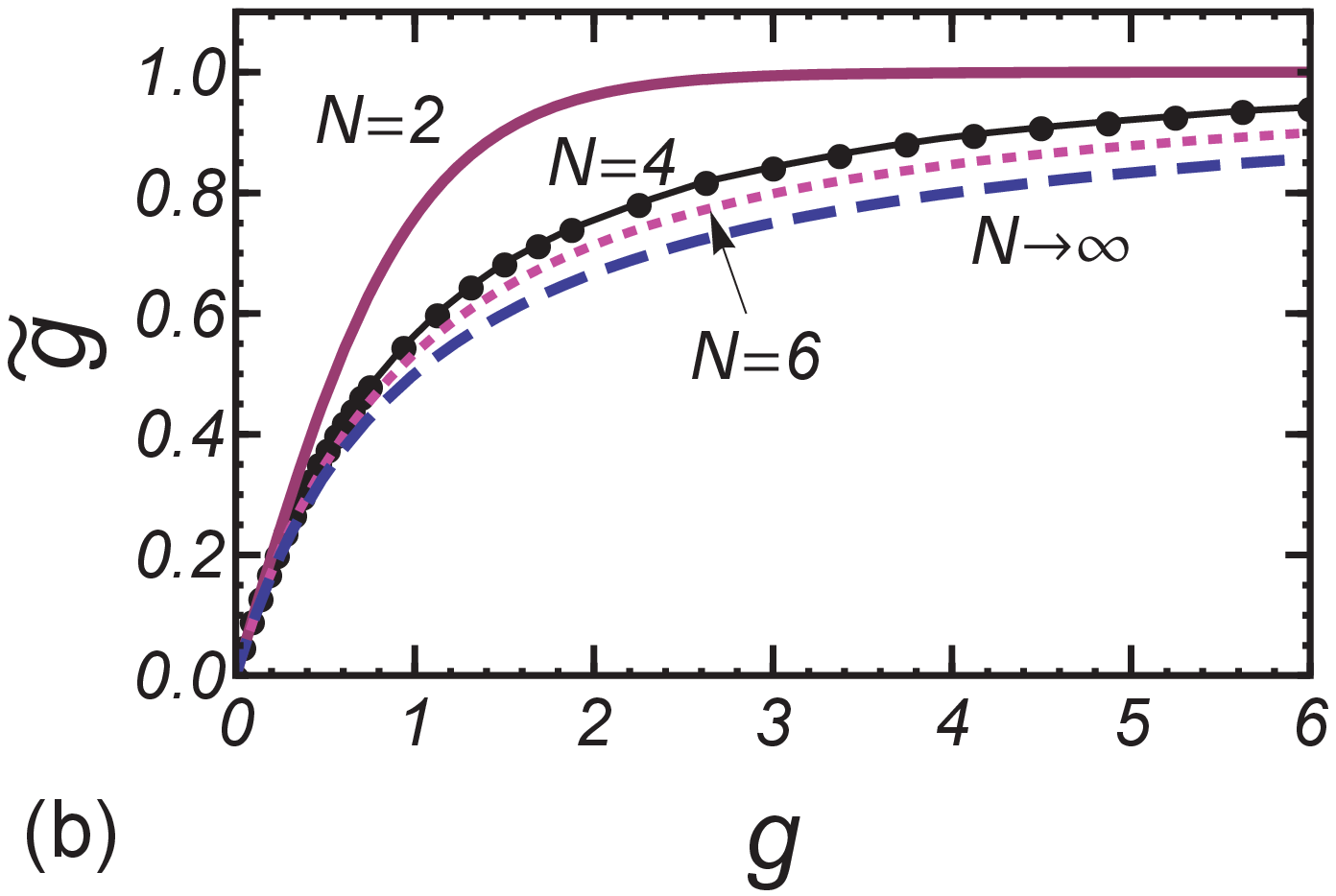}
\end{minipage}
\vspace{-0.6cm}
\caption{
(Color online) 
(a):  $\widetilde{g}$ and $z$ versus $g$ for $N=4$.
The curve with the circles represents the NRG results.   
The red dotted line represents the order $1/(N-1)$ results for $\widetilde{g}$,
and the blue dashed line the order $1/(N-1)^2$ results for $z$.
(b): $\,\widetilde{g}\,$ vs $g$ for $N=2$ 
(Bethe ansats \cite{ZlaticHorvatic}),
 $N=4$ (NRG), $N=6$ (order $1/(N-1)$), 
and for $N \to \infty$ where $\widetilde{g} \to g/(1+g)$. 
}
\label{fig:g*_g}
\end{figure}

To calculate the renormalized coupling constant $\widetilde{g}$ 
to order $1/(N-1)$,  we need $\Gamma_{mm';m'm}^{}(0,0;0,0)$ 
to order $1/(N-1)^2$ as $\widetilde{g}$ has a scaling factor $N-1$ 
defined in Eq.\ \eqref{eq:g_scale}.
The order $1/(N-1)^2$ contributions to the vertex function 
arise from the diagrams shown in Fig.\ \ref{fig:vertex_rpa}, 
and from the order $1/(N-1)^2$ component  
of the vertex diagram in Fig.\ \ref{fig:sg_rpa}. 
Summing up all these contributions, 
$\widetilde{g}$ can be expressed in the form 
that is exact up to terms of order $1/(N-1)$,
\begin{align}
\widetilde{g} = & \  
\frac{g}{1+g} 
\frac{1+ 
\frac{g}{N-1}
\left[1+\left(2 - \frac{g}{1+g}\right)
\mathcal{I}_{\phi}^{}(g)
\right]
}{1 +  \frac{g}{N-1} 
\left[\frac{g}{1+g} + \mathcal{I}_{\phi}^{}(g)\right]
}  
+ O\!\left(\!\frac{1}{N'^2}\!\right). 
\label{eq:vertex_rpa_correction}
\end{align}
Here, 
$\mathcal{I}_{\phi}^{}(g) \equiv \pi \Delta \int  
 \! \frac{d\omega}{2\pi} 
 \left\{G_0^{}(i\omega)\right\}^2\Pi(i\omega)$, 
and $\,N'\equiv N-1$. 
This formula shows the correct asymptotic form 
in both the weak and the strong coupling limits:    
$\widetilde{g} \simeq g$ for $g \to 0$, and 
$\widetilde{g} \to 1$  for $g \to \infty$.
Thus, Eq.\ \eqref{eq:vertex_rpa_correction}
can also be regarded as an interpolation formula for 
the Wilson ratio as $R-1=\widetilde{g}/(N-1)$ at half-filling.
The order $1/(N-1)$ results for $\widetilde{g}$ 
show an excellent agreement with the NRG results 
for $N = 4$ 
as indicated 
in Fig.\ \ref{fig:g*_g} (a).

To obtain Eq.\ \eqref{eq:vertex_rpa_correction}, 
the parameter $\widetilde{\gamma}$ in the denominator 
has been taken into account up to order $1/(N-1)$,   
\begin{align}
\widetilde{\gamma}
=& \ 
1 +  
\frac{g}{N-1} 
\left[
\frac{g}{1+g} 
+ 
\mathcal{I}_{\phi}^{}(g) 
\right]
+ \widetilde{\gamma}^{(\frac{1}{N'^2})}
+ O\!\left(\!\frac{1}{N'^3}\!\right)  . 
\end{align}
We also calculate, $\widetilde{\gamma}^{(\frac{1}{N'^2})}$,
the order $1/(N-1)^2$ contributions 
which arise from the diagrams shown in Fig.\ \ref{fig:sg_rpa_n2} 
and from the higher order component 
of the self-energy diagram in Fig.~\ref{fig:sg_rpa}.

Figure \ref{fig:g*_g} (a) shows a 
comparison between 
the NRG \cite{Sakano,Nishikawa1} and the $1/(N-1)$ expansion 
results for $N=4$. 
We see the very close agreement, 
especially  for $\widetilde{g}$.
Although the order $1/(N-1)$ results are slightly smaller 
than the NRG results,  the two curves for $\widetilde{g}$ 
almost overlap each other over the whole range of $g$.   
The deviation must decrease as $N$ increases.
Therefore, the order $1/(N-1)$ formula for $\widetilde{g}$ 
given in Eq.\ \eqref{eq:vertex_rpa_correction} 
provides almost exact numerical values for $N > 4$.
We also see in Fig.\ \ref{fig:g*_g} (b)   
the value that $\widetilde{g}$ can take  
is bounded in a very narrow region  
between the curve for $N=4$ and that for the $N\to \infty$ limit.  
As $N$ increases, $\widetilde{g}$ 
varies rapidly towards the value for the large $N$ limit.
The order $1/(N-1)^2$ results for the renormalization factor $z$, 
shown in Fig.\ \ref{fig:g*_g} (a), 
also agree with the NRG results for $N=4$ 
at $g \lesssim 3.0$, or equivalently $\widetilde{g} \lesssim 0.8$, 
from the weak to the intermediate coupling region 
 where $\widetilde{g}$ is still not converged to $1.0$,   
the value for the strong coupling limit.
Therefore, away from the strong coupling regime 
the Kondo energy scale, $\widetilde{\Delta} = z \Delta$,  
can be deduced reasonably from 
the order $1/(N-1)^2$ results.

\begin{figure}[h]
 \leavevmode
\begin{minipage}{0.49\linewidth}
\includegraphics[width=1\linewidth]{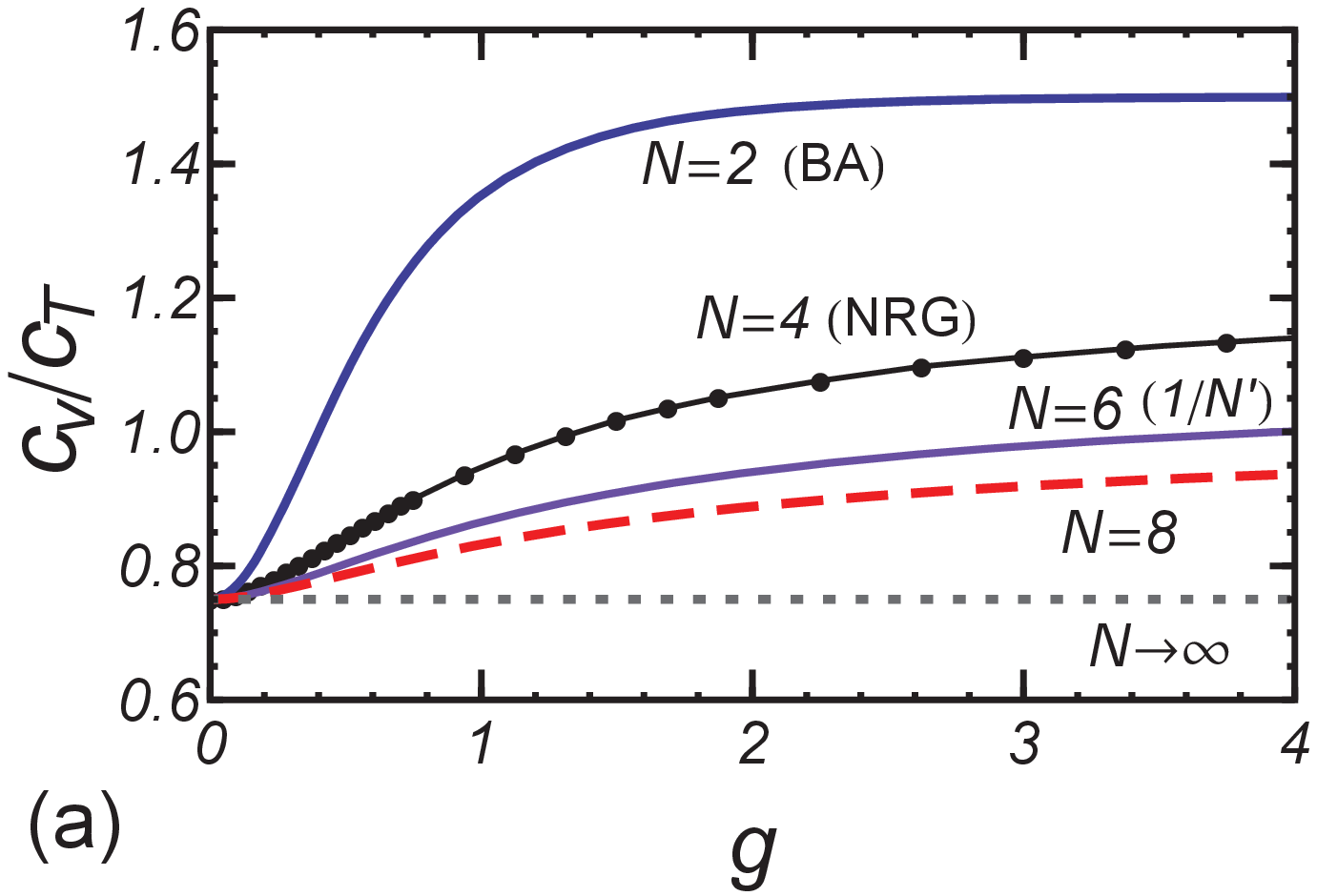}
\end{minipage}
\begin{minipage}{0.49\linewidth}
\includegraphics[width=1.0\linewidth]{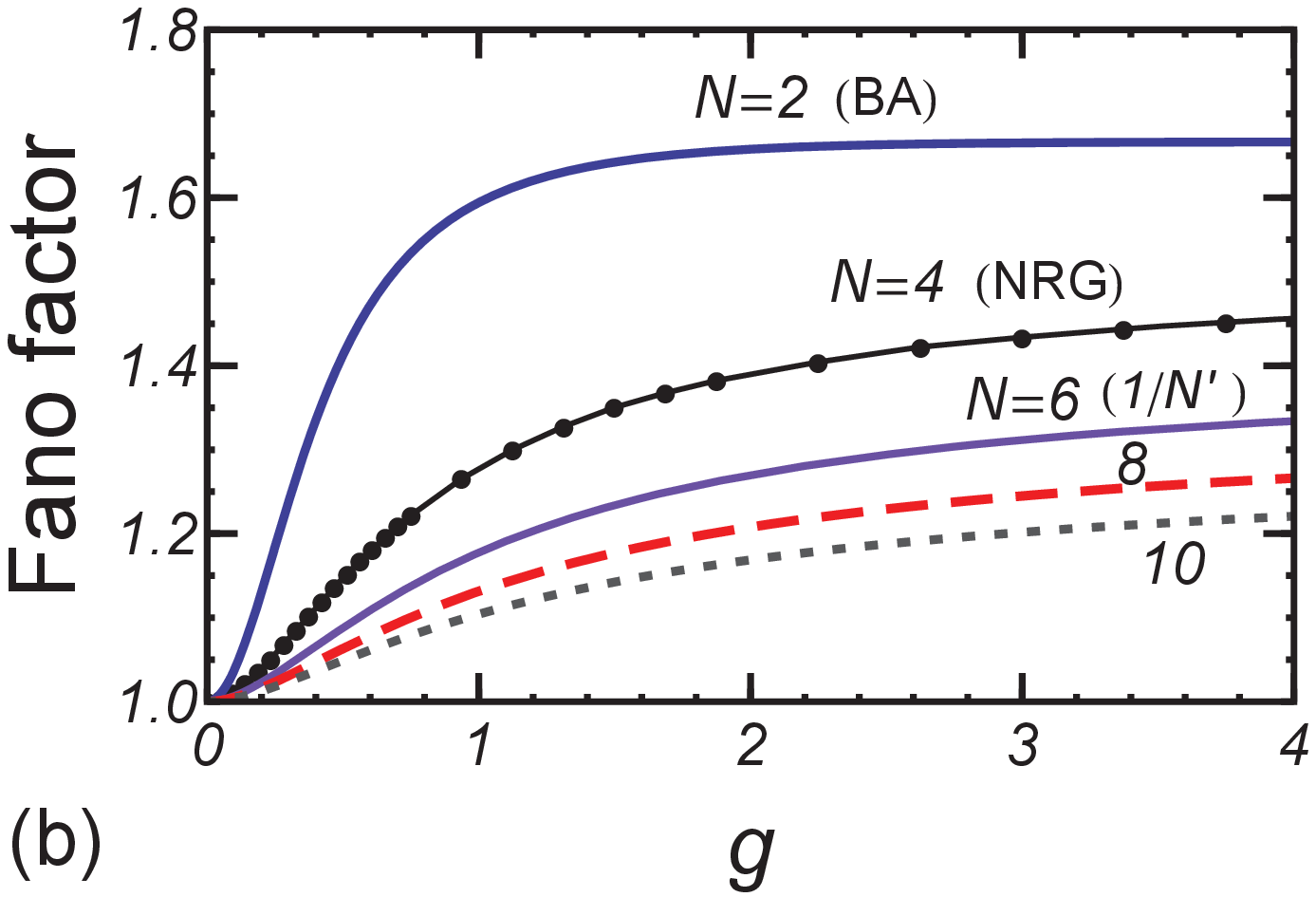}
\end{minipage}
\vspace{-0.2cm}
\caption{(Color online) 
Plots of (a)  $c_V/c_T$  and (b) $F_b\,$
as a function of $g$  for $N=2$ (Bethe ansats), $N=4$ (NRG), 
and for $N \geq 6$ the order $1/(N-1)$ results.
In the $N\to \infty$ limit, 
the curves approach to (a) $c_V/c_T \to 3/4$ and  (b) $F_b \to 1$. 
}
 \label{fig:cv_ct}
\end{figure}

The $1/(N-1)$ expansion can be applied fruitfully to 
nonequilibrium transport at finite $U$.
To be specific, we choose the lead-dot couplings and chemical potentials 
to be symmetric: $\Gamma_L = \Gamma_R$  and  $\mu_{L}= -\mu_{R}$ ($= eV/2$).
In this case, an exact expression can be derived for 
the retarded Green's function 
at low energies up to order $\omega^2$, $T^2$, 
and $(eV)^2$ \cite{ao2001,Sakano},     
\begin{align}
G^r(\omega) 
 \simeq  
\frac{z}{ 
\omega + i \widetilde{\Delta} 
+i
\frac{{\widetilde{g}}^2}{2(N-1)\widetilde{\Delta}} 
\left[ \omega^2 + \frac{3}{4}(eV)^2 + (\pi T)^2 \right] 
}.
\end{align}
The differential conductance for the current through 
the impurity can be deduced 
from $G^r(\omega)$, using  
the formula by Meir-Wingreen \cite{MW} and Hershfield \cite{HDW}, 
\begin{align}
&
\frac{dJ}{dV}   
= 
 \frac{Ne^2}{h}  \left[ 
1  -  c_T^{}
      \left( \frac{\pi T}{\widetilde{\Delta}}  \right)^2
   - 
       c_V^{}
         \left( \frac{eV}{\widetilde{\Delta}} \right)^2 
  + \cdots 
 \right],
\\
&c_T^{} 
=
\frac{1}{3}
\left(1+\frac{2\,{\widetilde{g}}^2}{N-1} \right)
, 
\quad \  
 c_V^{}   
 =   
\frac{1}{4}
\left(1+\frac{5\,{\widetilde{g}}^2}{N-1}\right).
\label{eq:cVcT}
\end{align}
The low-energy behavior is characterized 
by the two parameters, 
 $\widetilde{g}$ in the coefficients and 
$\widetilde{\Delta}$ the energy scale, which depend on $N$. 
Figure  \ref{fig:cv_ct} (a) shows the ratio of $c_V^{}$ to $ c_T^{}$  
as a function of $g$ for several $N$, 
using Eq.\ \eqref{eq:vertex_rpa_correction} for $N \geq 6$.
The ratio takes a value 
in the range ${3}/{4}  \leq c_V^{}/c_T^{} \leq (3/4)(N+4)/(N+1)$ 
\cite{FootNote3}.
The order $1/(N-1)$ results for $\widetilde{g}$ are numerically almost 
exact for $N>4$ 
as mentioned, and thus the results shown in Fig.\ \ref{fig:cv_ct}
capture orbital effects correctly.

As another application of Eq.\ \eqref{eq:vertex_rpa_correction}, 
we also consider the shot noise 
$S= 
\int 
  \! dt \,
\langle
\delta \hat{J}(t)\delta \hat{J}(0) +
\delta \hat{J}(0)\delta \hat{J}(t) \rangle$, 
where $\delta \hat{J}(t) \equiv 
\hat{J}(t) - \langle J \rangle$ is the current operator. 
At $T=0$, $S$ has been calculated 
to order $(eV)^3$ for the symmetric Anderson model 
for $N=2$ \cite{Sela2009,Fujii2010}, and  for general $N$:
$ S  
 =
 \frac{N e^2}{h}
 \frac{1}{6} \bigl(1 + \frac{9{\widetilde{g}}^2}{N-1}\bigr)
 \bigl( \frac{eV}{\widetilde{\Delta}} \bigr)^2     
  eV$ \cite{Sakano}.
The Fano factor $F_b$ is defined as the ratio of $S$ to
the backscattering current $J_b=NeV/h -J$, 
and has been obtained in the form \cite{Sakano},  
\begin{align}
F_b \equiv \frac{S}{2e J_b} \  = \ 
\frac{1+\frac{9\,{\widetilde{g}}^2}{N-1}}
{1+\frac{5\,{\widetilde{g}}^2}{N-1}} \;.
\label{eq:FanoFactor}
\end{align}
It takes a value in the range $1 \leq F_b \leq  (N+8)/(N+4)$.  
In Fig. \ref{fig:cv_ct} (b), 
the order $1/(N-1)$ results for $F_b$ are plotted 
as functions of $g$ for $N \geq 6$, 
together with the exact results for $N \leq 4$ \cite{Sakano}. 
As $N$ increases, $\widetilde{g}$ converges rapidly to 
the value, $\widetilde{g} \simeq g/(1+g)$,  for the large $N$ limit,
as mentioned in the above.
Thus, for $N \gtrsim 8$, the $N$ dependence is determined essentially 
by the factor $1/(N-1)$, seen explicitly in Eq.\ \eqref{eq:FanoFactor}.
The $1/(N-1)$ expansion can also be applied to    
the full counting statistics \cite{SakanoFCS}.

In conclusion, 
we have described the $1/(N-1)$ expansion approach 
based on the scaling defined in Eq.\ \eqref{eq:g_scale}.
The next leading order results for $\widetilde{g}$, 
which at half-filling corresponds to $\widetilde{g}=(N-1)(R-1)$,  
can be expressed in the form of 
Eq.\ \eqref{eq:vertex_rpa_correction}.
We find that 
this formula interpolates almost exactly between 
the weak  and the strong coupling limits 
for $N \geq 4$.
The $1/(N-1)$ expansion can be extended to explore 
the particle-hole asymmetric case \cite{FootNote1}.
Furthermore, it provides  a well-defined and controlled way 
to take into account the fluctuations 
near the $N \to \infty$ fixed point of  many fermion systems 
with two-body interactions.

The authors thank J.\ E.\ Han,  A.\ C.\ Hewson, 
and S.\ Tarucha for discussions.
This work is supported by 
the JSPS Grant-in-Aid for 
Scientific Research C (No.\ 23540375)  and S (No.\ 19104007).
Numerical computation was partly carried out 
at Yukawa Institute Computer Facility.




\begin{thebibliography}{99}



\bibitem{Hewson_book}
A.\ C.\ Hewson, 
 { \em  The Kondo Problem to Heavy Fermions\/} 
 (Cambridge University Press, Cambridge, 1993). 



\bibitem{Nozieres}
P. Nozi\`{e}res, J.\ Low Temp.\ Phys.\ {\bf 17}, 31 (1974).




\bibitem{YY2}
  K. Yamada, 
  Prog.\ Theor.\ Phys. {\bf 53}, 970 (1975).

\bibitem{ZlaticHorvatic}
V.\ Zlati\'c and B.\ Horvati\'c,
Phys.\ Rev.\ B {\bf 28}, 6904 (1983).


\bibitem{Yoshimori}
 A.\ Yoshimori, Prog.\ Theor.\ Phys.\ {\bf 55}, 67 (1976).





\bibitem{Grobis}
M.\ Grobis, 
I.\ G.\ Rau, R.\ M.\ Potok, H.\ Shtrikman, 
and D.\ Goldhaber-Gordon,
Phys.\ Rev.\ Lett.\ {\bf 100}, 246601 (2008).


\bibitem{ScottNatelson}
G.\ D.\ Scott, 
 Z.\ K.\ Keane, J.\ W.\ Ciszek, J.\ M.\ Tour, and D.\ Natelson, 
Phys.\ Rev.\ B {\bf 79}, 165413 (2009).






\bibitem{KNG}
A.\ Kaminski, Yu.\ V.\ Nazarov, and L.\ I.\ Glazman,
  Phys.\ Rev.\ B {\bf 62}, 8154 (2000).



\bibitem{ao2001}
A.\ Oguri, Phys.\ Rev.\ B {\bf 64}, 153305 (2001).


\bibitem{FujiiUeda} 
T.\ Fujii and K.\ Ueda, Phys.\ Rev.\ B {\bf 68}, 155310  (2003).



\bibitem{HBA} 
A.\ C.\ Hewson, J.\ Bauer, and A.\ Oguri,
J.\ Phys.: Condes.\ Matter.\  {\bf 17}, 5413 (2005).






\bibitem{Delattre} 
T.\ Delattre {\it et al}.,
 Nature Phys.\ {\bf 5}, 208 (2009).



\bibitem{GogolinKomnik} 
A.\ O.\ Gogolin and A.\ Komnik, 
 Phys.\ Rev.\ B {\bf 73}, 195301 (2006).

\bibitem{Golub} 
A.\ Golub, Phys.\ Rev.\ B {\bf 73}, 233310 (2006).


\bibitem{Mora2009}
C.\ Mora, P.\ Vitushinsky, X.\ Leyronas, A.\ A.\ Clerk, and
K.\ Le Hur, Phys. Rev. B {\bf 80}, 155322 (2009).

\bibitem{Sela2009} 
E.\ Sela and J.\ Malecki,
Phys.\ Rev.\ B {\bf 80}, 233103 (2009). 


\bibitem{Fujii2010}
T.\ Fujii, J.\ Phys.\ Soc.\ Jpn.\ {\bf 79}, 044714 (2010). 





\bibitem{Sakano}
R.\ Sakano, T.\ Fujii, and A.\ Oguri, 
Phys.\ Rev.\ B {\bf 83}, 075440  (2011).







\bibitem{Nishikawa1}
Y.\ Nishikawa, D.\ J.\ G.\ Crow, and A.\ C.\ Hewson, 
Phys.\ Rev.\ B {\bf 82}, 115123 (2010).


\bibitem{WilsonKogut} 
K.\ G.\ Wilson and J. Kogut, 
 Phys.\ Rep.\ C {\bf 12}, 75 (1974). 


\bibitem{Bickers}
N.\ Bickers, Rev.\ Mod.\ Phys.\ {\bf 59}, 845 (1987).


\bibitem{Haule}
K.\ Haule, S.\ Kirchner, J.\ Kroha, and P.\ W\"{o}lfle
Phys.\ Rev.\ B {\bf 64}, 155111 (2001).


\bibitem{OtsukiKuramoto}
J.\ Otsuki, and Y.\ Kuramoto
J.\ Phys.\ Soc.\ Jpn.\ {\bf 75}, 064707 (2006).






\bibitem{DMFT}
A.\ Georges, G.\ Kotliar, W.\ Krauth, and M.\ J.\ Rozenberg,
Rev.\ Mod.\ Phys.\ {\bf 68}, 13 (1996).





\bibitem{FootNote1} 
The Hartree type self-energy is included into  
$E_d \equiv \epsilon_d + \pi\Delta g\, \langle n_{dm}\rangle$. 
Thus, we find $\widetilde{g} \to g/[1+g/(1+(E_d/\Delta)^2)]$ 
in the large $N$ limit away from half-filling. 
 




 
\bibitem{MW} 
Y.\ Meir  and  N.\ S.\ Wingreen, 
Phys.\ Rev.\ Lett.\ {\bf 68}, 2512 (1992).


\bibitem{HDW}
 S.\ Hershfield, J.\ H.\ Davies, and J.\ W.\ Wilkins, 
 Phys.\ Rev.\ B {\bf 46}, 7046 (1992).



\bibitem{FootNote3}
In our definition,
the experimental value 
by Grobis {\it et al\/} \cite{Grobis} 
should be rescaled by a factor $\pi^2$ as $c_V^{}/c_T^{}=0.99 \pm 0.15$, 
and that of Scott {\it et al\/} \cite{ScottNatelson}
as $c_V^{}/c_T^{}=0.50 \pm 0.1$. 



\bibitem{SakanoFCS}
 R.\ Sakano, A.\ Oguri, T.\ Kato and S.\ Tarucha,
 Phys.\ Rev.\ B {\bf 83}, 241301 (2011).




\end{thebibliography}
\end{document}